\documentclass[prb,aps,twocolumn,floatfix]{revtex4}
\usepackage{graphicx}
\usepackage{amssymb,amsmath}

\newcommand{\av}[1]{\left\langle{#1}\right\rangle}

\begin{document}
\title{Decoherence and Full Counting Statistics in a  Mach-Zehnder   Interferometer}
\author{Heidi F\"orster$^1$, Sebastian Pilgram$^{1,2}$, and Markus B\"uttiker$^1$}
\affiliation{$^1$D\'epartement de Physique Th\'eorique,  Universit\'e de
  Gen\`eve,  CH-1211
  Gen\`eve, Switzerland\\
  $^2$Theoretische Physik, ETH Z\"urich, CH-8093 Z\"urich, Switzerland}
\date{\today}

\begin{abstract}
We investigate the Full Counting Statistics of an electrical Mach-Zehnder
interferometer penetrated by an Aharonov-Bohm flux, and in the presence of a
classical fluctuating potential.  Of interest is the suppression of the
Aharonov-Bohm oscillations in the distribution function of the transmitted
charge.  For a Gaussian fluctuating field we calculate the first
three cumulants.   The fluctuating potential causes
a modulation of the conductance leading 
in the third cumulant
to a term cubic in voltage and 
  to a contribution correlating modulation of current  and noise.   In the
  high voltage regime 
we present an approximation of the generating function. 
\end{abstract}
\maketitle
\section{Introduction}\label{sec:introduction}
For two decades theoretical and experimental investigations of the noise 
properties of small electrical conductors have been an important frontier of 
mesoscopic physics \cite{blanter2000}. Non-equilibrium noise, in particular 
shot noise \cite{khlus1987,lesovik1989,buttiker1990,buttiker1992}  
can provide information on physical properties which are not accessible from
conductance measurements.  
Recently there has been considerable interest in characterizing 
transport in mesoscopic systems not only via conductance and noise but also 
with all higher order current correlations. This is achieved by 
deriving a generating functional from which the quantities of interest 
can be obtained simply by taking derivatives. To construct the generating 
functional, it is useful to imagine that electrons passing the sample 
in a given time are counted, and the approach is thus known as Full Counting 
Statistics \cite{levitov1996, levitov1993}. While initial work was based
on the scattering approach to electrical transport, subsequently a number 
of different methods have been developed. These include an approach based
on Keldysh Green's functions \cite{nazarov1999,belzig2001a,belzig2001b,
  nazarov2002}, the non-linear sigma model\cite{gutman2004}, an
approach 
based on a cascade of Boltzmann-Langevin 
equations \cite{nagaev2002} and a formulation in terms of a stochastic 
path integral
\cite{pilgram2003,pilgram2004a,nagaev2004,pilgram2004b,jordan2004,jordan2004b, sukhorukov2004,bodineau2004,bertini2005}.                  
These later methods are principally useful if the quasi-classical part 
of the transport coefficients is dominant. 
A first experiment \cite{reulet2003} has given 
additional impetus for this research. It has stimulated work which treats 
the effect of the measurement circuit on the counting statistics
\cite{kindermann2004} and led to proposals for novel  
detection schemes \cite{tobiska2004,heikkila2004,pekola2004}.

In this work we are interested in the Full Counting Statistics (FCS) of
transport  
problems in which quantum effects are important. For instance,
in single channel mesoscopic conductors quantum interference is essential.
In such a situation, transport becomes classical only due to dephasing. It is 
thus interesting to examine the effect of inelastic processes on the counting 
statistics of a single channel problem and to investigate the transition 
from quantum mechanical transport to fully classical transport. 

\begin{figure}[t]
\begin{center}
	\includegraphics[width=\columnwidth]{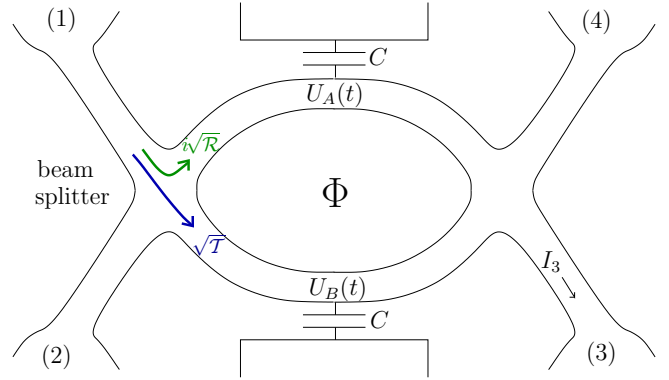}
	\caption{The Mach-Zehnder interferometer: a four-terminal AB-ring,
	threaded by a magnetic flux.   The interferometer arms are
	coupled to classical fluctuating potentials modeling the effect of
	dephasing.  The influence of dephasing on current fluctuations in lead
	$3$ is investigated.}\label{mzi} 
\end{center}
\end{figure}

The central aim of our work is thus the derivation of the generating
functional in the presence of inelastic scattering. 
We investigate an interferometer penetrated by an Aharonov-Bohm flux 
and subject to a random classical fluctuating potential. For simplicity 
we assume that the interferometer is of the 
Mach-Zehnder (MZI) type (Fig.~\ref{mzi}), an 
interferometer without backscattering. In such a system there are no 
closed orbits and thus 
only a finite number of possible trajectories.

Despite these simplifying assumptions the derivation of the generating
functional  
is a non-trivial problem. To investigate the effect of the fluctuating
potential, 
a time-dependent treatment of transport is needed together with 
a technique of statistical averaging. In fact, in the present work 
we are able to give the generating 
functional only in the limiting case of the high voltage regime. 
In the 
general case, we present results for the first three cumulants.

The model we investigate is closely related to earlier work by Seelig et
al. on the  
effect of dephasing on the conductance  \cite{seelig2001} and four-terminal
resistances  
in ballistic interferometers \cite{seeligconfig2003}. In few channel 
structures dephasing might be sample specific \cite{seeligconfig2003}. 
Experimental investigations on dephasing in ballistic interferometers is
provided by Hansen et al.\cite{hansen2001} and Kobayashi et
al.\cite{kobayashi2002}. 
Recently an electronic MZI has been experimentally realized by Ji et
al.\cite{ji2003}, using 
edge states in the quantum hall regime. This experiment also raised 
interesting questions on the shot noise of the interferometer and 
dephasing. 
The effect of dephasing 
on shot noise in such an interferometer has been discussed by Marquardt and
Bruder based
on classical\cite{marquardt2004a} 
and
quantum\cite{marquardt2004c} fluctuating field 
models.  

Still another motivation for the work presented here derives 
from recent proposals to implement 
orbital entanglement \cite{samuelsson2003} based on electron-hole generation
\cite{beenakker2003} 
in two-particle intensity-interferometers \cite{samuelsson2004}.
In these proposals quantum interference effects play an important role in
noise and higher order correlations\cite{beenakker2004}.

\section{The Mach-Zehnder interferometer} \label{sec:mzi}
The conceptually simplest system displaying electron interference is the
electronic Mach-Zehnder interferometer.  
The setup is shown in Fig.~\ref{mzi}, it consists of two beam
splitters connected by two interferometer arms enclosing a magnetic flux.
Each beam splitter is characterized by a transmission amplitude
$\sqrt{\mathcal{T}}$ 
for an electron going straight through, and a reflection amplitude
$i\sqrt{\mathcal{R}}$, with \mbox{$\mathcal{R}+\mathcal{T}=1$}.  The simplicity
of the interferometer originates from the 
exclusion of backscattering which discards closed orbits and separates
completely the processes of left and right moving particles.

Interference occurs, because the different vector potentials in the two arms
lead to a phase difference between the paths $A$ and $B$.  This difference
creates a characteristic flux-periodicity in the interference pattern, the
Aharonov-Bohm effect.  

In our work, dephasing is introduced with the help of 
classical fluctuating potentials, $U_A(t)$ and $U_B(t)$.  
The
correlation function of these fluctuating potentials can be self-consistently
determined treating electron-electron interactions in a random phase
approximation\cite{seelig2001}.  Here we take this
correlation function to be given.  
The potentials cause the particles to pick up an
additional time-dependent phase \mbox{$\varphi_i(t) =
  \int_t^{t+\tau}\!\!dt'U_i(t')$}, $i=A$ or $B$ (here and in the following we
set \mbox{$e=\hbar=1$}).  We assume, that the traversal
time $\tau$ of the particles is not affected by the potentials.  The
transmission amplitudes of particles entering the interferometer at time $t$
in lead $\beta$ and leaving it at time $t'$ in lead $\alpha$ are collected in
the time-dependent scattering matrix\cite{polianski2003}         
$S_{\alpha\beta}(t,t')$.  
For a one-channel MZI this scattering matrix is four dimensional and has
special properties:  as backscattering is excluded, the scattering matrix has
non-vanishing elements only in the off-diagonal $2\times 2$ blocks.  Because
the 
motion is purely ballistic, the scattering matrix depends effectively on one
single time \mbox{$S(t,t')=\delta(t'-t-\tau)S(t)$}, where
\begin{eqnarray}
	S_{31}(t)& = & i
	\sqrt{\mathcal{RT}}(e^{i\Phi + i\varphi(t)}+1)\label{Smatrix}\\
	S_{32}(t) & = & \mathcal{T} e^{i\Phi +i\varphi(t)}
	-\mathcal{R}   \nonumber\\
	S_{41}(t) & = & -\mathcal{R} e^{ i\Phi +i\varphi(t)}
	+\mathcal{T} \nonumber\\
	S_{42}(t) & = & i
	\sqrt{\mathcal{RT}}(e^{i\Phi + i\varphi(t)}+1).\nonumber
\end{eqnarray}
Here we assumed equal arm lengths and introduced the phase difference
\mbox{$\varphi(t) = 
\varphi_A(t)-\varphi_B(t)$} and the magnetic flux $\Phi$ in units of the flux
quantum. 
The remaining four elements are found from the magnetic field symmetry of the
scattering matrix 
\mbox{$S_{\alpha\beta}(\Phi) = S_{\beta\alpha}(-\Phi)$}.  

Without the fluctuating potentials the system is coherent:
\mbox{$\varphi(t)=0$}.  The transmission probability becomes
\mbox{$T_{31}=|S_{31}|^2 =2\mathcal{RT}(1+\cos\Phi)$} and shows cosine
oscillations in flux.  Such AB-oscillations are visible in
conductance experiments. 

When we include dephasing by the fluctuating potentials, the scattering matrix
becomes time-dependent and the quantities of interest have to be averaged over
the  
statistically distributed potential.  This average suppresses the
AB-oscillations.  For the case of conductance and noise, these 
effects have been studied in detail\cite{seelig2001,marquardt2004a}.  In this
work 
we will generalize this discussion to higher order current correlations (Full
Counting Statistics).

\section{FCS of the coherent system}\label{sec:fcs}
The Full Counting Statistics\cite{levitov1993} (FCS) is the distribution
function $P(Q)$  giving the probability that
 a certain charge 
\mbox{$Q=\int_0^{t_0}dtI(t)$} flows through a system during the observation
time $t_0$.  Its Fourier transform is
called its generating function 
$\chi(\lambda)$, the conjugated variable $\lambda$ is named the counting
field:
\begin{eqnarray}
	\chi(\lambda)   &=& \sum_Q P(Q) e^{i \lambda Q}\label{FTP}\\ 
	P(Q)		&=& \frac{1}{2\pi} \int_{-\pi}^{\pi}\!\!d\lambda
	\chi(\lambda)e^{-i \lambda Q}. \label{Pn}
\end{eqnarray}
The expansion of $\ln\chi(\lambda)$ in $\lambda$ yields all irreducible
cumulants: 
\mbox{$\ln\chi(\lambda)=\sum_k\frac{(i\lambda)^k}{k!}\av{Q^k}$}. For large
$t_0$  the first cumulant gives the mean current $\av{I} = \av{Q}/t_0$,  the
second cumulant is the zero frequency noise \mbox{$\Sigma =
  \av{Q^2}/t_0$},  the third cumulant
\mbox{$C^3 = \av{Q^3}/t_0$} describes the asymmetry (skewness) of the
distribution, and so 
on.  The generating function provides in principle a complete
description of the zero-frequency current fluctuations.  

We consider a four-terminal conductor, thus $Q$ and $\lambda$ are four
dimensional vectors.  Since in the MZI current conservation relates current
fluctuation 
of all leads, it is in most cases sufficient to concentrate on the
charge 
distribution in one lead.  In the following we will chose lead $3$ and call
$Q$ its charge and $\lambda$ the corresponding counting variable.  

In the coherent regime ($\varphi(t)=0$), the scattering matrix is time
independent, and the calculation of the generating function is straightforward.
Following the work of Levitov and Lesovik\cite{levitov1993},  
one defines an energy-dependent $4\times4$ matrix \mbox{$A =
  f\cdot(S^\dagger \tilde S-1)$}, where the diagonal matrix
    \mbox{$f=f(E)$} contains the Fermi functions of the four
    terminals. $\tilde S$ incorporates the counting fields by phase shifts
\mbox{$	\tilde S_{\alpha \beta} = e^{i(\lambda_\alpha-\lambda_\beta)}
	S_{\alpha \beta}$} (remember that 
    \mbox{$\lambda_\alpha =\lambda\delta_{\alpha3}$}).  
The generating function is
\begin{equation}\label{chi}
	\ln \chi( \lambda)=\frac{t_0}{2\pi} \int\!dE tr \ln(1+A). 
\end{equation}
Because backscattering is excluded, the matrix $A$ is block diagonal, and
$\ln\chi$ separates into two independent contributions.  This fact expresses
the statistical independence of left and right going particles.

\begin{figure}[t]
	\includegraphics[width=0.55\columnwidth]{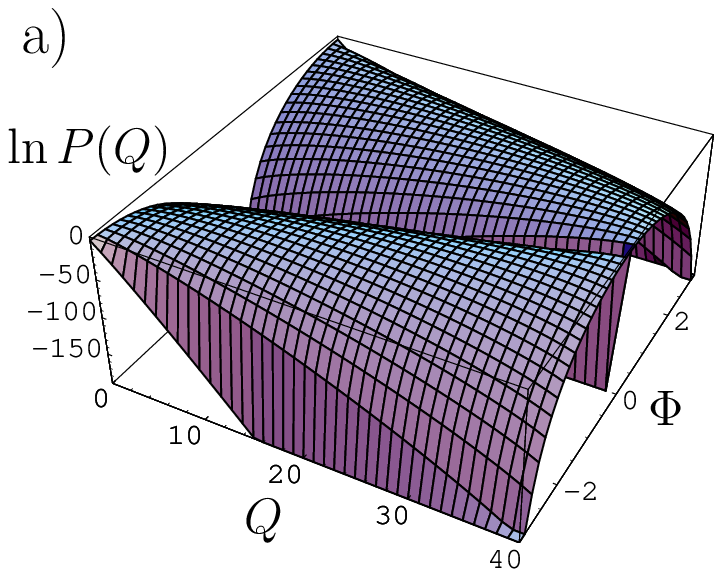} 
	\includegraphics[width=0.43\columnwidth]{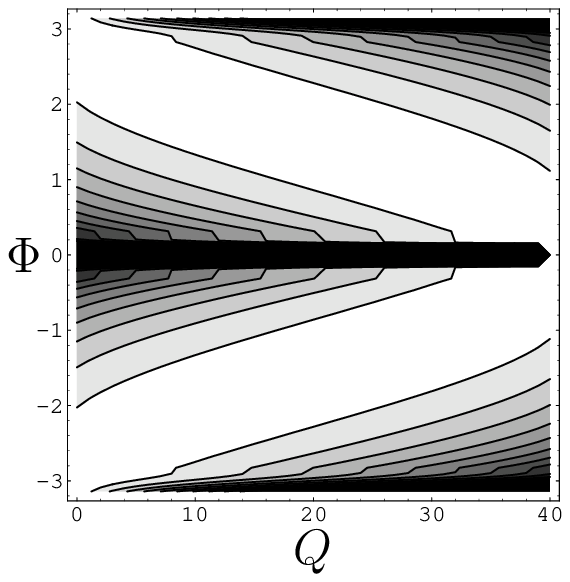} 
	\includegraphics[width=0.55\columnwidth]{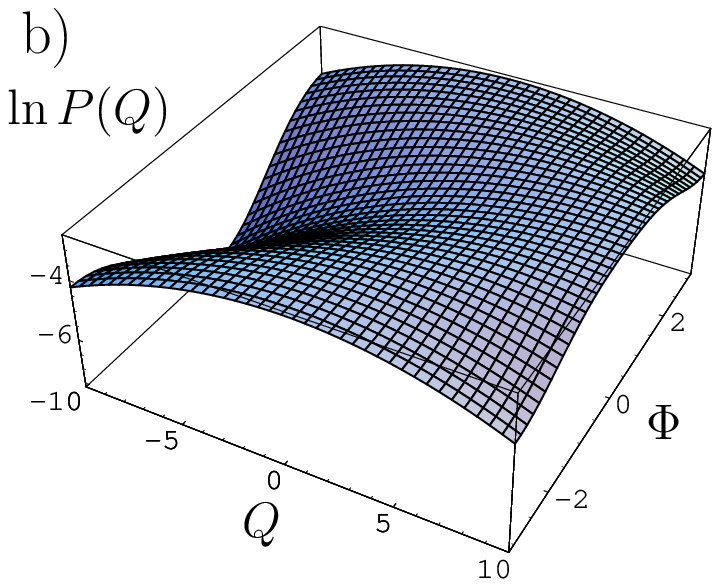} 
	\includegraphics[width=0.43\columnwidth]{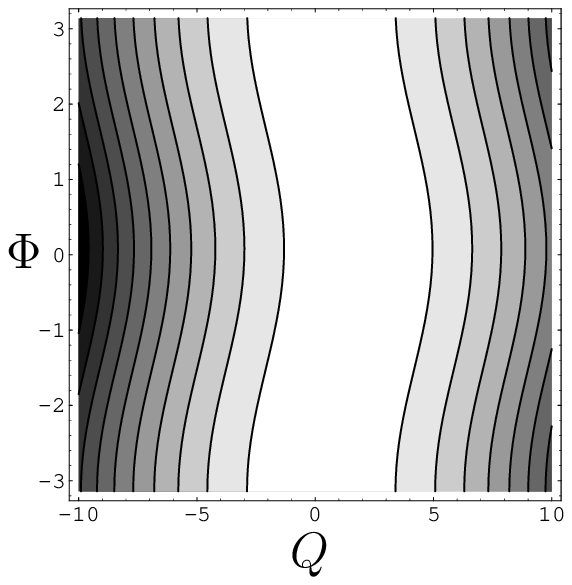} 
	\caption{(a) The distribution function at
	zero temperature 
	(logarithmic scale, $Vt_0/2\pi=40$).  Left: 3D-plot of the binomial
	distribution, which is periodic in flux. Right: the same in 
	contour representation, dark areas correspond to low probability. (b) 
	The distribution function at high temperature $k_BT=10 V$ (for a long
	observation time $t_0k_BT=100$). The flux dependence is washed out,
	and the distribution tends towards a Gaussian. }\label{P(Q)}
\end{figure}

For a two-terminal conductor, the FCS  can always be interpreted as a
classical binomial distribution. For a four terminal conductor like the MZI
this is in general no longer the case.  There exists however a special
configuration: at zero temperature a voltage $V$ is applied on terminal $1$ and
zero voltage on the other terminals.  In this case only two processes exist: 
transmission from terminal $1$ into lead $3$ with probability $T_{31}$ or
into lead $4$ with probability $1-T_{31}$.  The number of incident
particles during the time $t_0$  
is given by \mbox{$N=\frac{Vt_0}{2\pi}$}. 
We then recover the binomial distribution of the two-terminal conductor 
\begin{eqnarray}
	\chi(\lambda) &=&(e^{i\lambda}T_{31}+(1-T_{31}))^N\\
	P(Q) &=& {N \choose Q} T_{31}^{Q} (1-T_{31})^{N-Q}	
\end{eqnarray}
with the only difference, that the transmission probability
\mbox{$T_{31}=2\mathcal{RT}(1+\cos\Phi)$} depends 
on the magnetic flux.  Consequently the distribution function $P(Q)$ contains
all 
harmonics up to $\cos N\Phi$.  The flux periodicity of $P(Q)$ is shown in
Fig.~\ref{P(Q)}(a).
For symmetric beam splitters and certain values of $\Phi$, all particles end
up in one lead, due to complete constructive interference.  The distribution
is then singular and noiseless: 	\mbox{$P(Q)=\delta_{Q,N}$} for
\mbox{$\Phi=2\pi k$},  and
	\mbox{$P(Q)=\delta_{Q,0}$} for \mbox{$\Phi=(2k+1)\pi$} with
	\mbox{$k \in \mathbf{N}_0$}.    

At finite temperature this simple binomial picture is modified and does not
coincide with the two-terminal conductor statistics\cite{levitov1996}.  We
obtain 
\begin{equation}
	\chi(\lambda) =
	\exp\left[\frac{t_0k_BT}{2\pi} \left\{{\rm
	arcosh}^2u(\lambda)- v^2
	+iv\lambda-\frac{3}{4}\lambda^2\right\}\right]
\end{equation}
with \mbox{$v=\frac{V}{2k_BT}$} and the complex function \mbox{$
	u(\lambda) =	T_{31}\cosh\left(v+i\frac{\lambda}{2}\right)  +
	(1-T_{31})\cosh\left(v-i\frac{\lambda}{2}\right)$}.

Fig.~\ref{P(Q)}(b) shows plots of the logarithm of the
	distribution function for finite
	temperature. 
At high temperatures the flux dependence is strongly reduced, as is to be
	expected 
for a general interferometer.  But the MZI with arms of equal lengths
	investigated here has the special property that it is
described by energy-independent transmission probabilities. Thus the
	conductance  
does not suffer from energy averaging and exhibits full Aharonov-Bohm
	oscillations  
even at high temperatures.  However, the distribution function becomes
	classical, because the high-temperature fluctuations
of transmitted charge are dominated by two-particle processes, whose 
probability is flux independent and equal to one.

At very high temperatures $k_BT \gg V$, we find a Gaussian function \mbox{$P(Q)
  \sim e^{-(Q-\av{Q})^2/\Delta Q^2}$} with \mbox{$\Delta Q^2
  =\frac{\pi}{2t_0k_BT} $}, centered around the mean charge
  \mbox{$\av{Q}=\frac{t_0}{2\pi}T_{31}V$}.

\section{Influence of dephasing}\label{sec:dephasing}
We now start to consider the influence of the fluctuating potential on the
FCS.  The scattering matrix becomes time-dependent, and the electrons can be
scattered between different energies.  Following the idea
of Ivanov and Levitov\cite{ivanov1993}, the different energies can be viewed
as an infinite number of independent 
scattering channels.  Switching to time-representation
\mbox{$S(t,t')=\delta(t'-t-\tau)S(t)$}, this leads to a
generalization of formula~(\ref{chi}) 
for the generating function, where we extend  the usual trace ($tr$) to a trace
in time-space ($\mbox{Tr}$) including time integration:
\begin{equation}\label{lnchi}
	\ln \chi( \lambda)= \mbox{Tr} \ln(1+A) 
\end{equation}
All matrices are defined in analogy to
  section~\ref{sec:fcs}: The matrix $A$
  depends on two time-variables
  \mbox{$A(t', 
  t)=n(t'-t)(S^\dagger(t) \tilde S(t)-1)$}; the diagonal matrix $n$ contains
  the Fourier transforms of the Fermi functions 
in the four terminals 
 \mbox{ $n_\alpha = n_\alpha(t-t') = \int\frac{dE}{2\pi} f_\alpha(E) e^{i E
  (t-t')}$}.  
It can easily be checked, that 
  Eq.~(\ref{lnchi})   reduces to Eq.~(\ref{chi}), when $A$ depends only
  on time-differences (elastic scattering). 
 
To study the influence of dephasing on the generating
function, it has to be averaged statistically over
all possible 
realizations of the fluctuating potential: \mbox{$\av{\chi(\lambda)}_U =
  \av{e^{\mbox{Tr} \ln (1+A)}}_U$}.  For a Gaussian
distribution, this average is expressed by a functional integral
\begin{equation}\label{averaging}
	\Bigl\langle\chi(\lambda)\Bigr\rangle_U = \int \mathcal{D}U
	\chi(\lambda,\{U\})e^{-\frac{1}{2}\! 
	\int_0^{t_0}\!dt\!\!\int\! dt'U(t)K(t-t')U(t')}. 
\end{equation}
The kernel $K$ is defined via  \mbox{$K(t) = \int\frac{d\omega}{2\pi}
	\Sigma_U^{-1}(\omega) e^{-i\omega t}$}, with the spectral function of
	the 	potential \mbox{$
\Sigma_U(\omega) = \int dt\av{U(t)U(0)}e^{i\omega t}$}.  Only potential
	differences matter: \mbox{$U(t)=U_A(t)-U_B(t)$}. 

In general the averaging can not be performed analytically.
Therefore, in a first step, we expand
$\chi(\lambda)$  in the
counting field $\lambda$ and calculate the average of single cumulants,
crosschecking known results\cite{seelig2001, marquardt2004a} and presenting new
results for the third cumulant.  In a
second step,
we will  derive
directly an expression for $\chi(\lambda)$ in the case of a slowly fluctuating
potential. 

\section{Cumulants, exact and limiting cases} \label{sec:cumulants}
The effect of the averaging on
the cumulants is determined by three quantities: the correlation time
$\tau_{RC}$ of 
the fluctuating potential $U(t)$, the traversal time $\tau$, and a decoherence
parameter $z$. 

$\tau_{RC}$ is typically the $RC$-time, with $C$ the coupling capacitance,
and $R$ the charge relaxation resistance\cite{buttiker1993}; it provides a
cut-off frequency for 
energy exchanges between particles and fluctuating potential.  
The traversal time $\tau$ is the correlation time of the 
phase
$\varphi(t)$, because the electrons integrate the fluctuating potential $U(t)$
during
time $\tau$.  For \mbox{$\tau_{RC} \ll \tau$} the correlation function of the
phase 
is a triangle in time\cite{note3} \mbox{$ 
  \av{\varphi(t)\varphi(0)} = \Sigma_U\cdot(\tau-|t|)\,\Theta(\tau-|t|)$} with 
$\Sigma_U$ the spectral function of the potential at zero frequency.
This spectral function is given by \mbox{$\Sigma_U=\Delta \tilde
  U^2\tau_{RC}$}, 
where $\Delta \tilde U$ is the  mean fluctuation amplitude 
of the potential.  Because the relevant quantity in the MZI is 
the accumulated phase $\varphi(t)$, it is convenient to introduce 
the reduced mean fluctuation amplitude \mbox{$\Delta U =
  \sqrt{\frac{\tau_{RC}}{\tau}}\Delta \tilde U$}.  
The spectral function $\Sigma_U$  defines --for a Gaussian fluctuating field
$U(t)$-- the
decoherence parameter $z$ via\cite{seelig2001,seelig2003,marquardt2004a}:
\begin{equation}
	z = \exp(-\frac{\Sigma_U\tau}{2}).
\end{equation} 
The parameter $z$ indicates the dephasing strength; it varies between $0$ for
complete dephasing and $1$ in the absence of dephasing. 

We consider the configuration with $V$  applied on terminal~$1$, and
current fluctuations observed in lead~$3$.
The cumulant of $k$-th order is given by
\begin{equation}\label{Qk}
	\av{Q^k} =
	\left.\frac{1}{i^k}\frac{d^k\ln\av{\chi(\lambda)}_{U}}{d
	\lambda^k}\right|_{\lambda=0}.
\end{equation}
The calculation for the first cumulants is sketched in the 
appendix~\ref{app:cumulants}, here we will give 
only the main results and discuss the figures. 

The {\it mean current } in lead $3$ is linear in voltage, and
proportional to  the
transmission probability, averaged over the potential: 
\mbox{$\av{I_3} = \frac{1}{2\pi}2\mathcal{RT}(1+z\cos\Phi) V$}. In comparison
to the coherent system, the
AB-oscillations in the conductance are suppressed by the decoherence parameter
$z$, in agreement with the work of Seelig and B\"uttiker\cite{seelig2001}. 

\begin{figure}[t]
	\includegraphics[width=0.47\columnwidth]{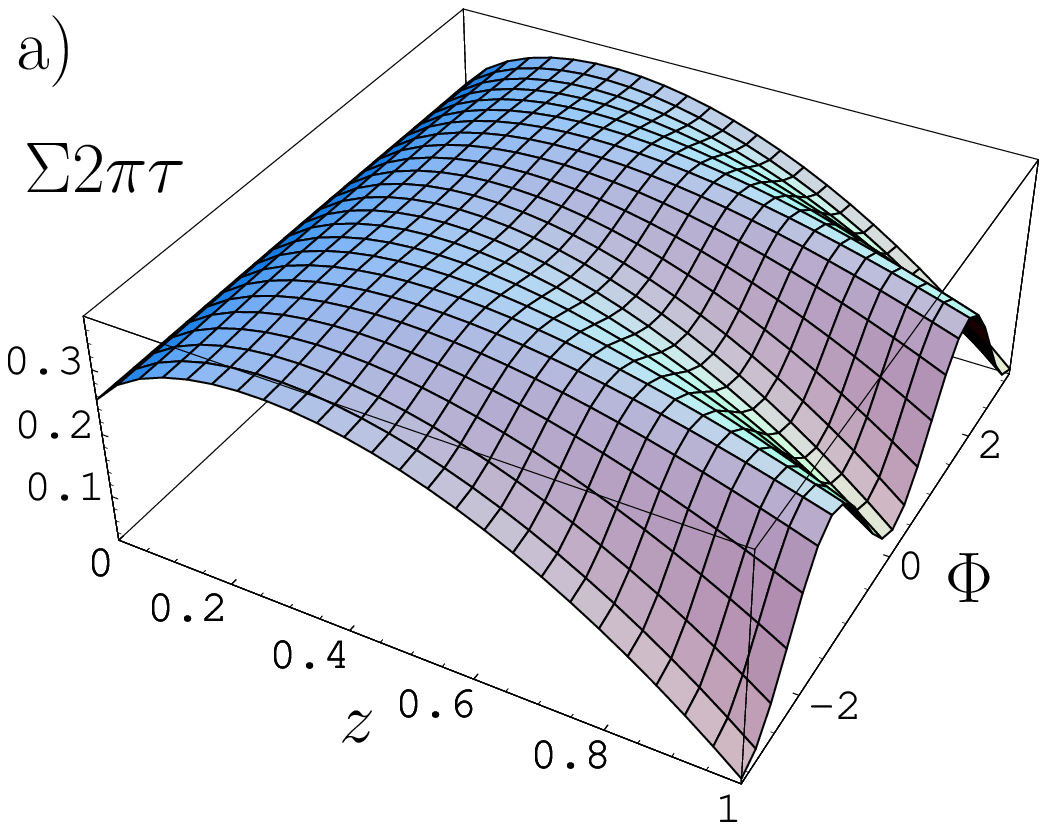} 
	\includegraphics[width=0.47\columnwidth]{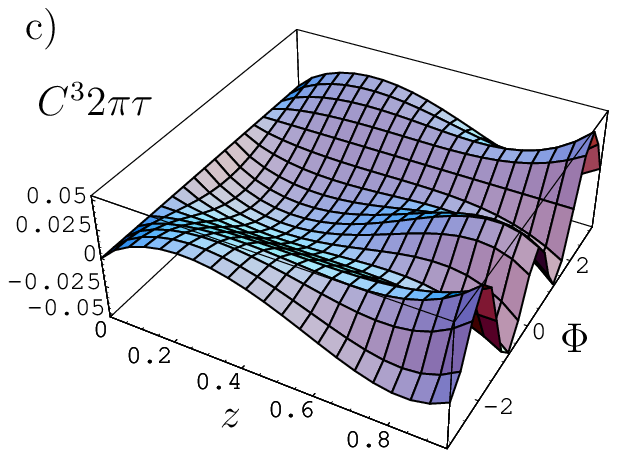}   
	\includegraphics[width=0.47\columnwidth]{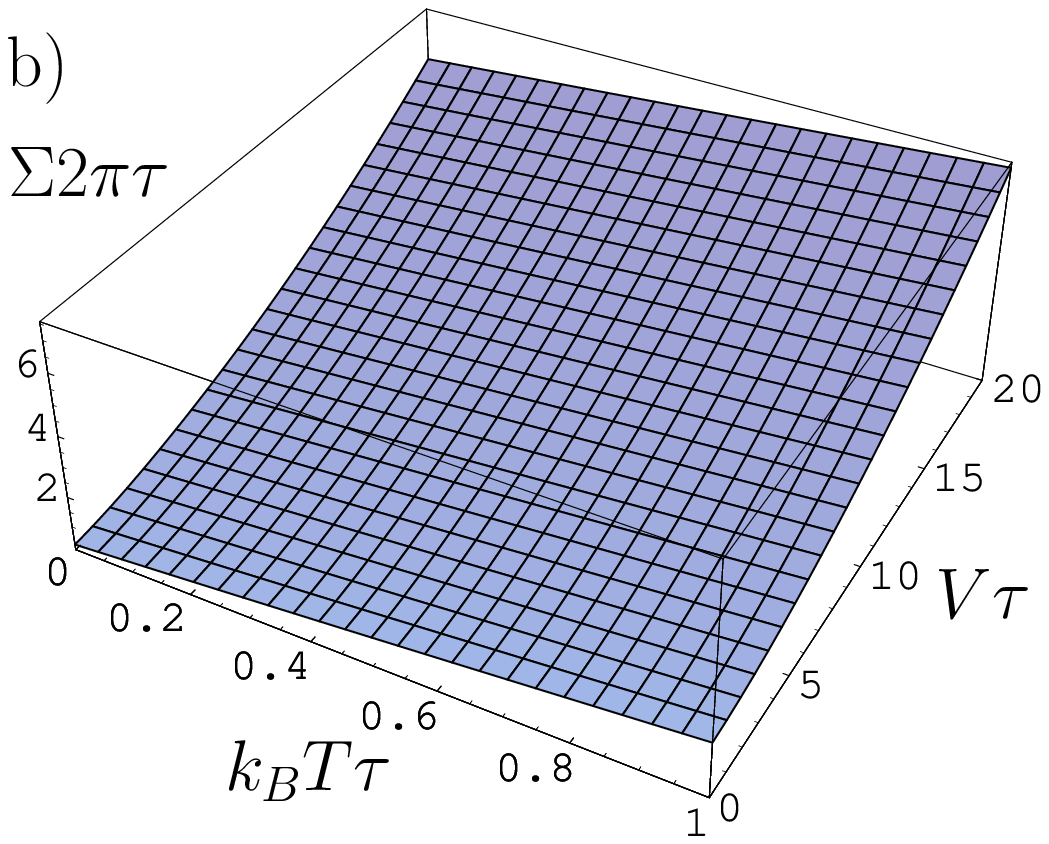}   
	\includegraphics[width=0.47\columnwidth]{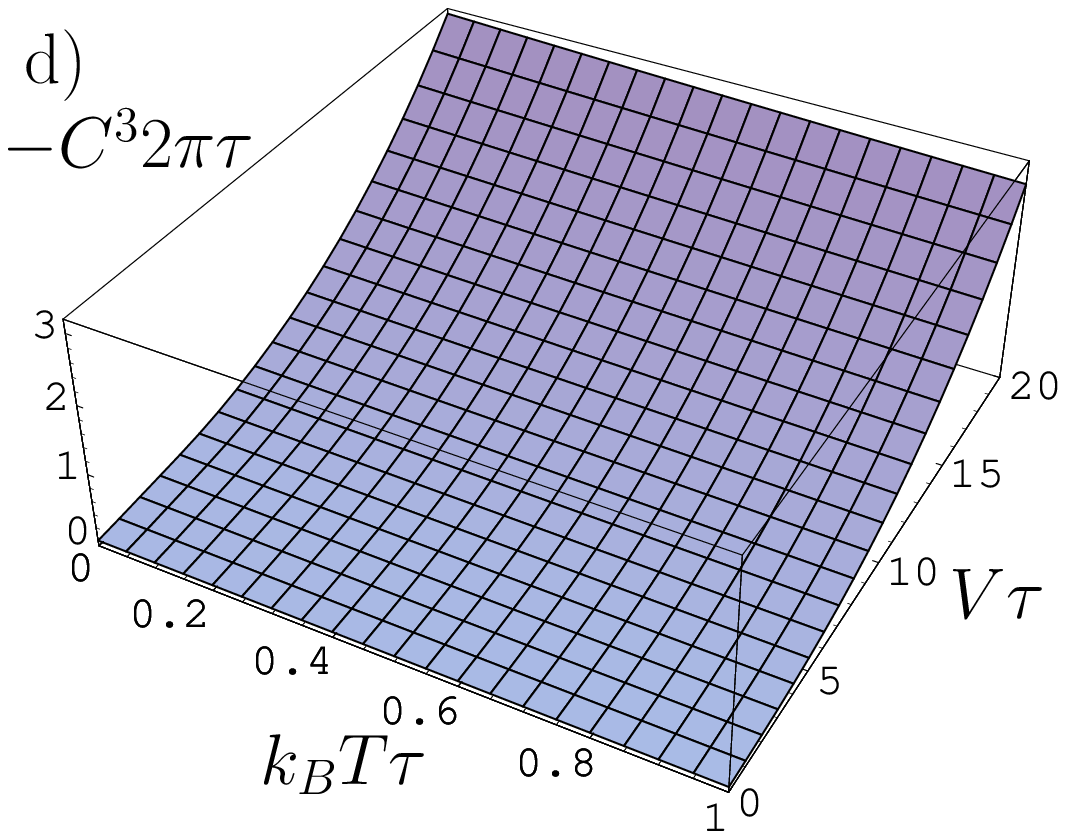}	
	\caption{Noise (a), ($k_BT=0$) and third cumulant (c), ($k_BT=0.1$) as
	a function of decoherence parameter $z$ and flux $\Phi$ (for
	$V\tau=1$): 
	the 	AB-oscillations are  suppressed for
	decreasing $z$.    (b) and (d) display the
	dependence on voltage and 
	temperature ($z=0.3$). 
	The third cumulant 
        changes little with temperature, but the noise shows
	linear thermal noise for high temperature.}\label{cumulants}
\end{figure}

The {\it noise} has been  studied in detail by Marquardt and
  Bruder\cite{marquardt2004a}.  
The expression for the noise is separated into a Nyquist part
  $\Sigma_{\text{nyq}}$, a modulation
part $\Sigma_{\text{I}}$ and  a two-particle exchange 
  contribution $\Sigma_{\text{ex}}$.   The $V$-independent Nyquist noise
  $\Sigma_{\text{nyq}}$ describes  thermal noise\cite{note1} as
  well as   
energy exchange between particles and the classical potential $U(t)$ up to the
  cut-off 
  frequency $1/\tau_{RC}$.  It is finite
  even for zero 
  temperature and voltage.  $\Sigma_{\text{I}}$ is 
a consequence of the variation of the
conductance in time,  it grows with $V^2$ and is
temperature-independent.  The exchange term  
$\Sigma_{\text{ex}}$ describes shot noise and is a function of voltage and
  temperature, vanishing for $V=0$ and linear for high voltage.
  $\Sigma_{\text{ex}}$ is a consequence of the indistinguishability of
  identical particles.  Thus the 
  total noise is 
\begin{equation}
	\Sigma =
	\Sigma_{\text{nyq}}(T)+\Sigma_{\text{I}}(V^2)+\Sigma_{\text{ex}}
	(V,T).
\end{equation}
For symmetric beam splitters, the fundamental AB-period is absent in the noise
and the AB-oscillations  are proportional to 
$\cos2\Phi$. They are  suppressed for $z\rightarrow 0$, compare
Fig.~\ref{cumulants}(a).   

In Fig.~\ref{cumulants}(b) the noise is shown as a
function of  temperature and
voltage: it is linear in  temperature  for
high $k_BT\tau$, simply due to thermal noise. The quadratic dependence on
voltage does not change with temperature. 

\begin{figure}[t]
	\begin{center}
	\includegraphics[width=0.9\columnwidth]{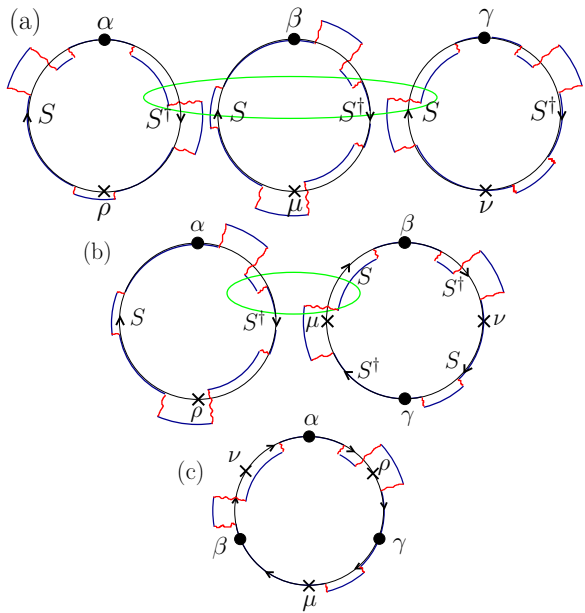}

	\vspace*{-3mm}
	\caption{Processes contributing to the third cumulant at zero
	temperature. The black
	points 
	correspond to channels $\alpha$, $\beta$, $\gamma$ at time $t+\tau$,
	and the crosses  to 
	channels $\rho$, $\mu$, $\nu$ at time t.  
	In  $C^3_{\text{I}}$ (a) three one-particle processes are connected by
	correlations of the fluctuating potential (circled in grey/green). In
	(b) the correlation of the modulated current and the shot noise are
	displayed, which contributes to $C^3_{\text{I}\Sigma}$. 
	$C^3_{\text{3p}}$ (c) is the three particle contribution, averaged over
	excitations.}\label{sketches}
	\end{center}
\end{figure}

In the {\it third cumulant} we distinguish three contributions: a modulation
part $C^3_{\text{I}}$, 
proportional to $V^3$, a second term $C^3_{\text{I}\Sigma}$, that connects
current modulation 
and noise, and a three-particle contribution
$C^3_{\text{3p}}$:  
\begin{equation}
	C^3 =
	C^3_{\text{I}}(V^3)+C^3_{\text{I}\Sigma}(V,T)+C^3_{\text{3p}}(V,T) 
\end{equation}
The processes leading to these three contributions can be visualized as
shown in Fig.~\ref{sketches}.  Fig.~\ref{sketches}(a) corresponds to the
modulation contribution $C^3_{\text{I}}$.  Each circle contains the 
one-particle process creating current in the corresponding lead:  
In the left half of the first circle an electron,
entering the scatterer at time $t$ in lead  $\rho$ is transmitted into lead
$\alpha$ at time $t+\tau$ with the 
amplitude $S_{\alpha\rho}$, it is moving forward in time.  Correspondingly in
the right half circle a hole is transmitted
from lead $\alpha$ into lead $\rho$ with amplitude $S^\dagger_{\rho\alpha}$,
moving  
backward in time.  We sum over all channels at the points
\includegraphics[width=0.7em]{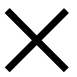} 
and measure the current at the points 
\includegraphics[width=0.7em]{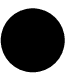}. As we investigate the third
cumulant of charge transmitted into lead $3$, we have here
\mbox{$\alpha=\beta=\gamma=3$}.   

The radial
direction can be interpreted as the energy of the 
particles.  Due to the fluctuating potential, we allow energy exchange and
average over all excitations during the traversal time $\tau$ under the
condition, that the energy takes the value $E$ at point
\includegraphics[width=0.7em]{point.eps}. 
 
The modulation part $C^3_{\text{I}}$ consists of three one-particle
processes, that are 
connected via the correlation of the 
fluctuating
potential only.

In the second term $C^3_{\text{I}\Sigma}$ a modulated current and the noise are
correlated also via the fluctuating potential.  
In the shot
noise (the right ring in Fig.~\ref{sketches}(b)) we have two points
\includegraphics[width=0.8em]{cross.eps} (time $t$)  and
\includegraphics[width=0.7em]{point.eps} (time $t+\tau$), 
because we deal with a two-particle process: 
an
electron propagates from channel $\mu$ to $\beta$, a hole from $\beta$ to
$\nu$, an 
electron from $\nu$ to $\gamma$, and a hole from $\gamma$ to $\mu$.  In
addition there is --even at zero temperature-- a correlation between the
modulated current and the Nyquist 
noise, which is not represented in Fig.~\ref{sketches}.

$C^3_{\text{3p}}$ then is the actual three-particle exchange contribution,
connecting 
the leads $\alpha$, $\beta$, and $\gamma$ by six $S$-matrices.  
Here we again average over 
excitations between the points \includegraphics[width=0.7em]{point.eps}.

\begin{figure}[t]
	\includegraphics[width=0.9\columnwidth]{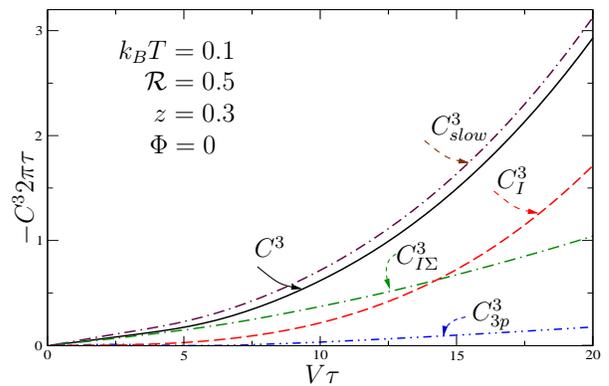} 
\caption{The third cumulant and its three contributions as
	a function of voltage. 
   At high $V\tau$ the cubic term dominates.  The dashed line
   $C^3_{\text{slow}}$ shows the approximation in the high voltage
	regime.}\label{cumV} 
\end{figure}

In Fig.~\ref{cumV} the three
contributions and the total third cumulant $C^3$ are depicted as a function of
$V\tau$. Note that $C^3$ is 
negative for $\Phi=0$ and $z=0.3$, this implies that the distribution in lead
$3$ is 
skewed to the right.  For high voltage
the term $C^3_{\text{3p}}$ is linear, $C^3_{\text{I}\Sigma}$  is quadratic,
and the cubic contribution  $C^3_{\text{I}}$ dominates. The full third 
cumulant is odd in voltage and thus vanishes for $V=0$.

In the symmetric case,
$\mathcal{R} = 0.5$, we find AB-oscillations proportional to  $\cos3\Phi$ and
$\cos\Phi$; 
with decreasing $z$ the terms in $\cos3\Phi$ are stronger damped than those
with $\cos\Phi$, as shown in 
Fig.~\ref{cumulants}(c).  Note also, that around \mbox{$\Phi=0,\pm\pi$} the
third 
cumulant changes sign with decreasing $z$. 
Fig.~\ref{cumulants}(d) shows the third cumulant as a function of 
temperature  and voltage. 
   In contrast to the noise the third cumulant
contains no voltage independent (Nyquist) contribution whatsoever, and
therefore, changes little with temperature\cite{levitov2004}.   

We can distinguish two limiting regimes: 
{\it high} and
{\it low} voltage\cite{marquardt2004a} in comparison to $1/\tau$.  In the
first case, many particles (\mbox{$V\tau \gg 1$})
pick up approximately the same phase
$\varphi(t)$, since many wave packets 
fit at the same time into an interferometer arm.  To the contrary 
for low voltage (\mbox{$V\tau \ll 1$}) every wave packet acquires
different phases $\varphi(t)$, which  are uncorrelated.  
Therefore, for high and low voltage we can make approximations of a slowly and
a fast varying phase $\varphi(t)$ respectively.

We next compare these approximations 
with the exact expression.
The modulation contribution $C^3_{\text{I}}$ does not depend on the phase
dynamics, but is determined by the parameter $z$.
Therefore, it is independent of whether we consider a fast or a slowly
varying phase. 
As 
$C^3_{\text{I}}$ decreases with  $V^3\tau^3$, in the low voltage regime it is 
small 
compared to the terms linear in voltage.
In both limiting cases the averaging of the contributions $C^3_{\text{3p}}$
and $C^3_{\text{I}\Sigma}$ simplifies;   additional
details of the averaging and the approximations are discussed in the
appendix~\ref{app:cumulants}. 

  The low voltage 
approximation fits only for very small values of $V\tau$, and we will not
discuss this limiting case here further.  A comparison of the high voltage
approximation, detailed in the appendix~\ref{app:cumulants}, with the exact
result 
is shown in Fig.~\ref{cumV}.  This approximation matches the exact curve very
well for a wide range of $V\tau$  and even in the case  \mbox{$V\tau \approx
  1$}.

\section{Generating function in the high voltage regime} 
The discussion above showed that in principle we can calculate any
single  cumulant with the
only assumption of a Gaussian fluctuating potential. However the calculational
effort grows very fast with the order of the cumulant.  For this reason let us
now 
concentrate on the limiting case of high voltage 
(\mbox{$V\tau \gg 1$}), when many particles 
pick up approximately the same phase.  
In this
limit  the generating function, Eq.~(\ref{lnchi}) can be evaluated directly. To
this end, we perform the averaging, see Eq.~(\ref{averaging}) under the
following two assumptions. 

(i) When the accumulated phase \mbox{$\varphi(t)=\int_t^{t+\tau}dt'U(t')$}
  varies slowly, the matrix
  $A(t,t')$ in Eq.~(\ref{lnchi}) takes the form\cite{note2} \mbox{$
  A(t,t')    \rightarrow n(t-t')(S^\dagger(\varphi)\tilde
  S(\varphi)-1)$}. It
  depends on a time difference and on a parameter $\varphi$, that
  itself is a function of a (single) time variable.   
After Fourier transformation we are left with one
  integral in time and another in energy. At zero temperature, the energy
  integral is easily performed and one obtains
\begin{equation}\label{chislow}
	\mbox{Tr}\ln(1+A) \rightarrow 
   \frac{V}{2\pi}\int_0^{t_0}\!\!dt\ln[1+T_{31}(\varphi(t))(e^{i\lambda}-1)]
 \end{equation}
Here $T_{31}(\varphi(t))$ depends on the instantaneous phase $\varphi(t)$. 

(ii) To take into account the correlation time $\tau$ of phase fluctuations,
we make an additional approximation beyond the slow approximation discussed so
far: We take the slowly varying phase $\varphi(t)$ to be a steplike function,
that has
constant values \mbox{$\varphi=\mathcal{U}\tau$} in time intervals
$[k\tau,(k+1)\tau]$ for \mbox{$k \in \mathbf{N}_0$}. 
The values $\varphi$ and consequently the values $\mathcal{U}$ obey a Gaussian
distribution. 
Note that the parameter $\mathcal{U}$ does not describe the actual
fluctuations of the potential 
$U(t)$ since these are determined by the correlation time $\tau_{RC}$ which is
here assumed to be smaller than $\tau$.

The $t_0/\tau$ time intervals of length
$\tau$ that can be treated
independently.  During each time interval the functional integral
(\ref{averaging}) 
becomes a simple Gaussian integral in the variable $\mathcal{U}$: 
\begin{equation}
	\Bigl\langle\chi(\lambda)\Bigr\rangle_U = \left(
	\int\!\frac{d\mathcal{U}e^{
	-\frac{\mathcal{U}^2}{2\Delta\!U^2}}}{\sqrt{2\pi \Delta\! 
	U^2}} (1+T_{31}(\mathcal{U})(e^{i\lambda}-1))^{N_\tau} 
	\right)^{\!\!\frac{t_0}{\tau}},\label{chidiscrete}
\end{equation}
with \mbox{$T_{31}(\mathcal{U}) = 2\mathcal{RT}(1+\cos(\Phi+\mathcal{U}\tau))$}.  
The integration in $\mathcal{U}$ leads to the generating function
\begin{equation}\label{chislowstrong}
  \Bigl\langle\chi(\lambda)\Bigr\rangle_U  =\!\left(\!\sum_{m=0}^{N_{\tau}}
	\!\sum_{p=0}^m\!{N_{\tau} \choose m} {m \choose p} c^{N_{\tau}-m} a^p
	b^{m-p} z^{(2p-m)^2}\!\!\right)^{\!\!\!\frac{t_0}{\tau}}
\end{equation}
with the constants \mbox{$c=
	1+2\mathcal{RT}(e^{i\lambda}-1)$},
	\mbox{$a=\mathcal{RT}e^{i\Phi}(e^{i\lambda}-1)$}, \mbox{$
	b=\mathcal{RT}e^{-i\Phi}(e^{i\lambda}-1)$}.

This approximation for the generating function applies also in the case
	\mbox{$\tau_{RC} > \tau$}: then the correlation time of the phase
	$\varphi(t)$ is the same as the one of the potential $U(t)$,
	$\tau_{RC}$, and the time $\tau$ in Eq.~(\ref{chislowstrong}) has to be
	replaced by $\tau_{RC}$.
\begin{figure}[t]
	\includegraphics[width=0.55\columnwidth]{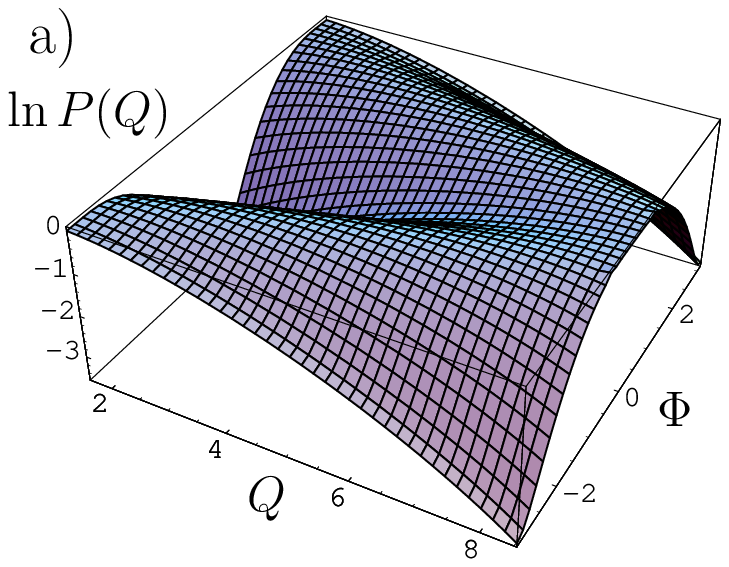} 
	\includegraphics[width=0.43\columnwidth]{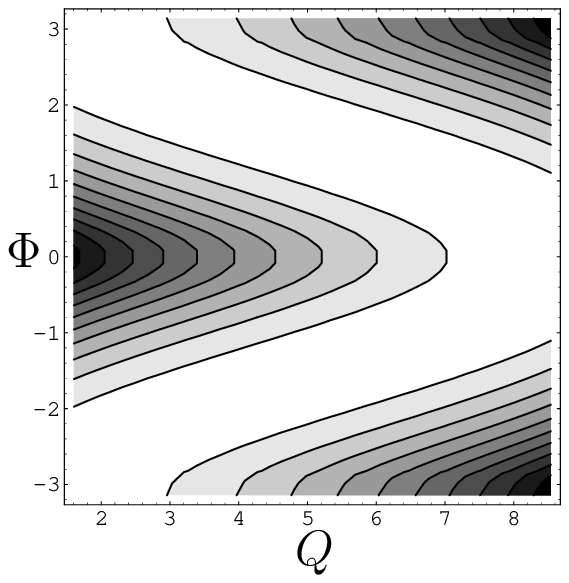} 
	\includegraphics[width=0.55\columnwidth]{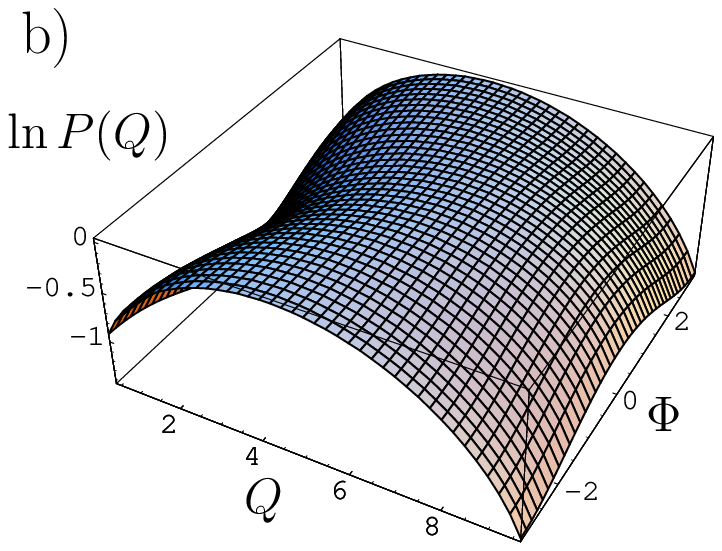}
	\includegraphics[width=0.43\columnwidth]{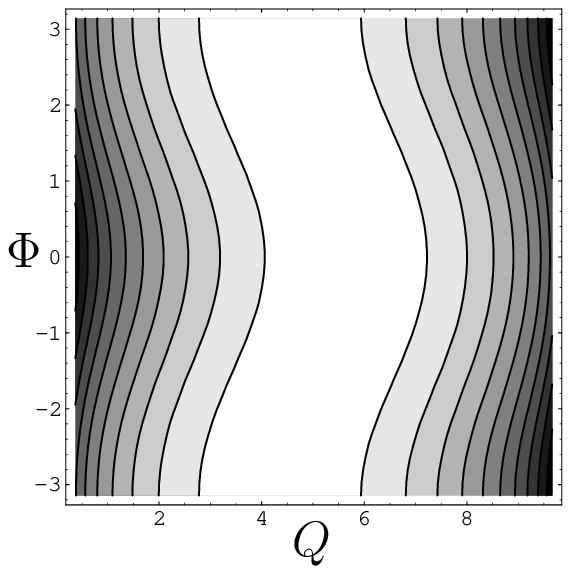}
	\caption{The distribution function (logarithmic scale) at zero
	temperature under influence of decoherence (\mbox{$N_\tau=10$},
	\mbox{$t_0=\tau$}).
	(a) weak
	dephasing: $z=e^{-0.32}$, 
	(b) strong dephasing: $z=e^{-2}$. The flux dependence is suppressed,
	and the distribution tends towards a symmetric 
	function for  $\mathcal{R}=0.5$.  }\label{PQdeph}
\end{figure} 

Starting from this generating function~(\ref{chislowstrong}) one finds
immediately the 
cumulants by simply taking derivatives in $\lambda$. 
These cumulants in their modulation contributions from
the slow limiting case of the cumulants presented in
section~\ref{sec:cumulants},  due to the treatment of $\varphi(t)$ as a
steplike 
function (corresponding to a discretization in time space). 

Each cumulant shows harmonic
oscillations in $\Phi$ up to the order of the derivative. As expected the
terms containing oscillations with
higher  periodicity  are damped stronger. 
Therefore, Eq.~(\ref{chislowstrong}) is a convenient starting point to
investigate the strong dephasing limit.  In this case we can restrict the sum
in Eq.~(\ref{chislowstrong}) to small powers in $z$. 
Up to first order in $z$ the generating function consists of two
hypergeometric functions $F$
\begin{eqnarray}
	\av{\chi(\lambda)}_U	& = & \left[ c^{N_{\tau}}
	F\!\left(\!-\frac{N_{\tau}}{2},-\frac{N_{\tau}}{2}+\frac{1}{2};1;\frac{ab}  
	{4c^2}\!\right)\right. +\\
	& & \hspace*{-2cm} \left.+
	c^{N_{\tau}-1}\!N_{\tau}(c-\!1)z\cos\Phi
	F\!\!\left(\!-\frac{N_{\tau}}{2}+\frac{1}{2},-\frac{N_{\tau}}{2}+1;2;\frac{
	ab}{4c^2}\!\right)\right]^{\!\frac{t_0}{\tau}}\!\!\!.\nonumber
\end{eqnarray}
The first term represents the purely classical part, it does not depend on $z$
and $\Phi$ but only on the beam
splitter parameter $\mathcal{R}$, $\lambda$ and $N_\tau$.  For
\mbox{$\mathcal{R}=0.5$} it is 
symmetric in $\lambda$ and thus yields a symmetric contribution to the
distribution function.  The second part is the interference term proportional
to $z\cos\Phi$; for \mbox{$\mathcal{R}=0.5$} it is asymmetric in~$\lambda$. 

In the general case for arbitrary $z$, we perform the Fourier transform of
Eq.~(\ref{chidiscrete}) numerically  to 
find  the distribution function 
$P(Q)$.  Fig.\ \ref{PQdeph} depicts the logarithm of the distribution
function for weak (Fig.~\ref{PQdeph}(a)) and strong dephasing
(Fig.~\ref{PQdeph}(b)) for a time interval \mbox{$t_0=\tau$}.  In
the case of weak dephasing, it  still exhibits
the form of a binomial distribution with the periodicity in $\Phi$
(compare with the no-dephasing case in Fig.~\ref{P(Q)}(a)).  For
increasing dephasing strength, the AB-effect is suppressed. In the
case of symmetric beam splitters the function $P(Q)$ then tends
towards 
a symmetric distribution around $N_\tau/2$, and the third cumulant vanishes. 
However, for asymmetric beam splitters the distribution function is not
symmetric. 

To extract information about the tails of the distribution function $P(Q)$,  
we use the stationary phase approximation and solve the saddle point equation
\mbox{$\frac{d\ln\av{\chi}}{d\lambda} = iQ$} for $\lambda$ in the complex
plane.  For \mbox{$Q \ll \av{Q}$}, i.e. when nearly no particles are
transfered to lead $3$, we find \mbox{$(-\lambda) \gg 1$} for negative
$\lambda$.
Starting from Eq.~(\ref{chislowstrong}) we obtain 
\begin{equation}\label{tail}
	\ln P(Q) \approx -Q\ln(
	\frac{\tau}{t_0}Q) +\frac{t_0}{\tau} \ln v-Q\ln\frac{v}{w}    
\end{equation}
The first term determines essentially the variation of the distribution
function with $Q$.  Only the parameter $v$ and $w$ are functions of
the magnetic flux and the decoherence parameter $z$:
\begin{eqnarray}
	v &=& \sum_{m=0}^{N_\tau}\!\sum_{p,q=0}^m 
	v_{mpq}\\
	w &=& \sum_{m=0}^{N_\tau}\sum_{p,q=0}^m
	v_{mpq}\frac{N_\tau-m+q}{1-q}\frac{2\mathcal{RT}}{1-2\mathcal{RT}}
	\nonumber   
\end{eqnarray}
with 
\begin{eqnarray}
 	v_{mpq} &=& {\!N_{\tau}\! \choose \!m\!}\!\!{\!m\! \choose
	\!p\!}\!\!{\!N_{\tau}\!\!-\!\!m\! \choose \!-q\!}\!\!{\!m\! \choose
	  \!q\!} \cdot\\
	&&\hspace*{-5mm}\cdot
	2^{-q}(\!-\!\mathcal{RT}\!)^{m\!-\!q}(\!1-2\mathcal{RT}\!)^{\!N_\tau\!-\!m\!+\!q}    
	e^{i\Phi(2p-m)}z^{(2p-m)^2}. \nonumber
\end{eqnarray}
Thus in Eq.~(\ref{tail}) the dominating contribution of the $Q$-dependence of
$P(Q)$ is classical, and the 
last two terms represent the coherent part. 
For positive \mbox{$\lambda \gg 1$}, the large $Q$ tail, \mbox{$Q\gg \av{Q}$},
of the distribution function is obtained in an analogous procedure.

\section{Conclusion} \label{sec:discuss}

We analyze the Full Counting Statistics of a mesoscopic, electrical
Mach-Zehnder 
interferometer penetrated by an Aharonov-Bohm flux, and in the presence of a
classical fluctuating potential. The Mach-Zehnder interferometer has the 
advantage that there are no closed orbits. Interference is only a consequence
of superposition of amplitudes in the out-going channels.    
Of interest
is the generating function of transfered charge for such a quantum coherent
conductor and the transition from coherent transport to classical transport. 

First, we discuss the probability distribution for a fully coherent 
Mach-Zehnder interferometer. At zero temperature the distribution of
transfered charge is binomial  
like in a two terminal conductor, but with a flux-dependent transmission
probability.  At finite temperature, the 
fact that we deal with a multi-terminal conductor leads to a different
distribution which
becomes classical due to flux-independent two-particle processes. 

In the presence of a fluctuating potential, assuming that it is  
Gaussian, we obtain the first
three cumulants as a function of the Aharonov-Bohm flux and a dephasing
parameter.  
The classical fluctuating potential
modulates the conductance. As a consequence the noise contains a term
proportional to voltage squared.  The same effect results in a term cubic in
voltage in the third cumulant.  In the third cumulant the conductance
modulation is correlated as well with the shot noise.  
The cumulant 
of the $k$-th order shows oscillations in the flux up to $\cos k\Phi$. As one
expects  
the high-order harmonics of the Aharonov-Bohm oscillations are suppressed much
stronger with increasing  dephasing than the low 
order oscillations. 
 
For the limiting case of high voltage 
we find the generating function. This permits, like 
in the coherent case, the calculation of the distribution 
function of the transfered charge. Therefore, in
this limiting case, we are able to follow the 
suppression of the AB-oscillations in the distribution function.

For the tails of the distribution function we find that they 
consist of two parts: a purely classical part, independent of 
dephasing, which determines 
the dependence on transfered charge and a quantum part which 
governs the flux dependence. 

The analysis of the conceptually simple case of a Mach-Zehnder
interferometer will hopefully provide a useful point of reference 
for future theoretical investigations of Counting Statistics in
the presence of quantum fluctuations. However, 
the Mach-Zehnder interferometer is not only of theoretical interest,
but recently has been realized experimentally\cite{ji2003}.  
The questions addressed here are thus also of interest for future 
experiments.

\begin{acknowledgments}
We thank Eugene Sukhorukov for fruitful discussions.
This work was supported by the Swiss National Science Foundation and the
program for Materials with Novel Electronic Properties. 
\end{acknowledgments}

\appendix*

\section{Calculation of single cumulants}\label{app:cumulants}
Successive derivatives in $\lambda$ of the logarithm 
of the
generating function, see Eq.~(\ref{Qk}), yield the cumulants. Using the
expression~(\ref{lnchi}) each cumulant is expressed in terms of derivatives
of the matrix $A$, introduced in section~\ref{sec:dephasing}.
Derivatives of $A$ of $k$-th order can be written in terms of a matrix 
\mbox{$K = S^\dagger P
  S$} and the 
matrix $n$, defined in section~\ref{sec:dephasing}.  $P$ is a projection
matrix with only non-vanishing element, \mbox{$P_{33}=1$}.  Note that $K$
has non-vanishing elements only in the upper diagonal $2\times 
2$ block.   
Because backscattering is excluded in the MZI, the product $PK$ vanishes and
we obtain the simple form 
\begin{equation}\label{dA}
	\left.\frac{d^kA}{d\lambda^k}\right|_{\lambda=0}=i^k
	n\left(K+(-1)^kP\right).
\end{equation}
The matrix $K$ is separated into constant matrices $\av{K}$ and $L$, and
time-dependent scalar 
functions as follows: \mbox{$K =
  \av{K}+L(e^{i\varphi(t)}-z)+L^\dagger (e^{-i\varphi(t)}-z)$}.  Here $z$ is
the decoherence 
parameter \mbox{$z = \av{e^{\pm i\varphi(t)}}_U$}, introduced in
section~\ref{sec:cumulants}.  The average is calculated by replacing the
generating function in 
Eq.~(\ref{averaging}) by the phase factor; at this point we explicitly assume
a Gaussian fluctuating potential $U(t)$, and use $\tau_{RC}\ll\tau$. 

The average of the matrix $K$ is denoted by 
$\av{K}$, it is determined only by the parameters $\mathcal{R}$,
$\Phi$, and $z$.  
The
averaging of the cumulants only 
affects the time-dependent functions ($e^{\pm i\varphi(t)}-z$): 
Depending on the order $k$ of
the cumulants  one obtains products
of the functions $(e^{\pm i\varphi(t)}-z)$ at $k$ different times.  The
averaging then 
defines correlation functions of different order.  

In the noise we get two
correlation functions, depending on one time variable and determined by the
parameter~$z$:  
\begin{eqnarray}
	g_{\pm}(t) &=& \frac{1}{z^2} \left \langle(e^{i\varphi(t)}-z)(e^{\pm
	i\varphi(0)}-z)\right \rangle_U =\nonumber\\
	&=&[z^{\pm2(1-\frac{|t|}{\tau})}-1]\Theta(\tau-|t|).\label{gpm}
\end{eqnarray}
The procedure of the average is similar to the calculation of the parameter
$z$. 

For the third cumulant   four correlations functions
$h_{+\pm\pm}$  
depending on two time variables are possible, for example 
\begin{equation}\label{h}
	h_{+-+}(t,t') =
	\frac{1}{z^{3}}\av{(e^{i\varphi(t)}-z)(e^{-i\varphi(t')}-z)(e^{i\varphi(0)}-z)}. 
\end{equation}
The functions $h(t,t')$ can be expressed in terms of
$g_{\pm}$~given by Eq.~(\ref{gpm}).

With Eqs.~(\ref{Qk}) and~(\ref{dA}) all cumulants take  a compact
form. 
The mean current in lead $3$  is
\begin{equation}
	\av{I_3}= \frac{1}{it_0} \left\langle\mbox{Tr}
    \left.\frac{d A}{d \lambda}\right|_{\lambda=0}\right\rangle_U =
    \frac{1}{t_0} \mbox{Tr}[n\av{K}-nP],
\end{equation}
leading to the result presented in section~\ref{sec:cumulants}.

For the  noise we express the second cumulant in terms of derivatives of $A$,
taken at \mbox{$\lambda=0$}:
\begin{eqnarray}
	\Sigma &=&\frac{1}{i^2t_0}
	\left \langle \left(  \mbox{Tr} \left.\frac{dA}{d\lambda}\right. - \left
	\langle \mbox{Tr} \left.\frac{dA}{d\lambda}\right. \right
	\rangle_U\right)^2 \right \rangle_U +\label{sigma}\\
	&&+ 
	\frac{1}{i^2t_0}\left \langle 
	\left.\mbox{Tr}\frac{d^2A}{d\lambda^2}\right.
	-
	\left.\mbox{Tr}
	\left(\frac{dA}{d\lambda}\right)^2 \right. \right \rangle_U\nonumber \,.
\end{eqnarray}
The first term corresponds to the modulation noise $\Sigma_{\text{I}}$ and
disappears in the 
case of no dephasing. The second term  is the two-particle contribution (that
separates into $\Sigma_{\text{nyq}}$ and $\Sigma_{\text{ex}}$ as explained
in section~\ref{sec:cumulants}).  Introducing the 
$K$-matrix, Eq.~(\ref{sigma}) simplifies to
\begin{eqnarray}
	\Sigma &=&\frac{1}{t_0}
	\left \langle \left(  \mbox{Tr}[nK-n\av{K}]\right)^2 \right \rangle_U
	+\\ &&+
	\frac{1}{t_0}\Bigl\langle\mbox{Tr}[nK(1-n)K]\Bigr\rangle_U,\nonumber  
\end{eqnarray}
since the product $K(1-K)$ vanishes.  
The final result contains the Fourier transforms of
the functions $g_{\pm}$ and a frequency integral, as also found in 
Ref.~[33]. 

As for the current and the noise, we
express the third cumulant in terms of derivatives of $A$ at
\mbox{$\lambda=0$}:
\begin{eqnarray}
	C^3 &=&\frac{1}{i^3t_0}\left[
	\av{\left(\mbox{Tr}\frac{dA}{d\lambda}-\av{\mbox{Tr}\frac{dA}{d\lambda}}\right)^3}  
	+\right.\label{Q3}\\
	&+&\!\!3
	\av{\!\!\left(\mbox{Tr}\frac{dA}{d\lambda}\!-\!\av{\mbox{Tr}\frac{dA}{d\lambda}}\right)\!\!
	\left(\mbox{Tr}\frac{d^2A}{d\lambda^2} 
	\!-\!\mbox{Tr}\left(\frac{dA}{d\lambda}\right)^2\!\right)\!\!} +\nonumber \\
	&+&\!\!\left.\av{\mbox{Tr}\frac{d^3A}{d\lambda^3}} + 2
	\av{\mbox{Tr}\left(\frac{dA}{d\lambda}\right)^3\!} - 3
	\av{\mbox{Tr}\frac{dA}{d\lambda}\,\frac{d^2A}{d\lambda^2}}\right] \nonumber.
\end{eqnarray}
The first term corresponds to the modulation contribution $C^3_{\text{I}}$,
the second term to the term $C^3_{\text{I}\Sigma}$, 
and the remaining part gives the three-particle contribution
$C^3_{\text{3p}}$.  One checks immediately, 
that the first two terms vanish in the absence of dephasing.  Using again the
compact expressions in $K$, Eq.~(\ref{Q3}) becomes 
\begin{eqnarray}
	C^3 &=&\frac{1}{t_0}\left[
	\av{\left(\mbox{Tr}[nK-n\av{K}]\right)^3}  
	+\right.\label{C3}\\
	&& +3\;
	\Bigl\langle\left(\mbox{Tr}[nK-n\av{K}]\right)
	\left(\mbox{Tr}[nK(1-n)K]\right)\Bigr\rangle +\nonumber \\
	&&+\;\;\Bigl.\Bigl\langle\mbox{Tr}[nK(1-n)K(1-2n)K]\Bigr\rangle\Bigr]
	\nonumber. 
\end{eqnarray}
The evaluation of Eq.~(\ref{C3}) is quite lengthy.  Introducing Fourier
transforms of the correlation functions $h$ and $g$ and of the occupation
functions $n$ describing the leads, one is left with an integral over two
frequencies.   This integral contains a convolution of $h$ or $g$ with
temperature and voltage dependent kernels.  We evaluate the integration
numerically. 

The limiting cases were introduced in
section~\ref{sec:cumulants}.  In the high voltage regime, the accumulated
phase varies slowly. 
Therefore, we evaluate the 
correlation functions $g_{\pm}$ and $h$~(see Eqs.~(\ref{gpm}) and (\ref{h}))
at equal times:  
\begin{equation}
  \begin{array}{lcrcl}
	\text{slow: } & & g_{\pm}(t) &\approx& g_{\pm}(0)\\
		      & & h(t,t')&\approx& h(0,0).\end{array}
\end{equation}
In this limit the correlation functions are entirely determined by the
parameter $z$. 
The frequency integrals in the final expression
of $C^3_{\text{3p}}$ are given by integrands at zero frequency and thus
simplify substantially.
As the modulation term $C^3_{\text{I}}$ does not depend on the phase dynamics,
it does not change in this approximation.  The
term $C^3_{\text{I}\Sigma}$ combines a modulation part and a two-particle
process.  Here we use \mbox{$h(t,t')  \approx h(0,t')$}.

The high voltage approximation does not catch contributions to the third
cumulant due 
to particle-hole excitations created by the fluctuating potential $U(t)$.
Similar to Ref.~[33], 
this leads to an unimportant
linear offset between the exact result and the high voltage result. For
better comparison, this offset is subtracted in Fig.~\ref{cumV}, which shows
a good agreement of the high voltage 
approximation with the exact result.

In the opposite case, for low voltage, 
the phase
factors in formula~(\ref{gpm}) and~(\ref{h}) at different times are
uncorrelated and thus can be averaged independently and vanish.
In this case, we use in $C^3_{\text{3p}}$
the approximation 
\begin{equation}
  \begin{array}{lcrcl}
	\text{fast: } & & g_{\pm}(t) &\approx& 0 \\
		      & & h(t,t') & \approx & 0.  \end{array}
\end{equation}
$C^3_{\text{I}}$  is proportional to $V^3\tau^3$
and thus in the low voltage regime small
compared to the terms linear in voltage.  For low temperature also the
contribution $C^3_{\text{I}\Sigma}$ is negligible.



\begin{thebibliography}{99}
\bibitem{note3}
  The edges of the triangle are rounded on the time scale~$\tau_{RC}$.
\bibitem{note2}
  The phase $\varphi(t)$ consists of a slowly varying part and a fast
  fluctuating contribution. The later is neglected here.  The fast
  fluctuating   term  leads to 
  particle-hole excitations which are for example responsible for the Nyquist
  part in the noise.  Therefore, the generating function in this
  limiting case does not account for any Nyquist contributions to the
  cumulants.
\end{thebibliography}
\end{document}